\documentclass[aps,twocolumn,floatfix, showkeys, superscriptaddress, nofootinbib]{revtex4}
\usepackage{amsmath}
\usepackage{graphicx,bm,hhline}
\usepackage{multirow}
\DeclareMathAlphabet{\pazocal}{OMS}{zplm}{m}{n}

\newcommand{\br}{{\bm r}}
\newcommand{\brp}{{\br}^\prime}

\newcommand{\wig}[1]{\mathrel{\hbox{\hbox to 0pt{\lower.6ex\hbox{$\sim$}\hss}\raise.4ex\hbox{$#1$}}}}

\makeatletter
\renewcommand*\env@matrix[1][\arraystretch]{%
  \edef\arraystretch{#1}%
  \hskip -\arraycolsep
  \let\@ifnextchar\new@ifnextchar
  \array{*\c@MaxMatrixCols c}}
\makeatother

\begin{document}
\title{Model for the electrical conductivity in dense plasma mixtures}

\author{C. E. Starrett}
\email{starrett@lanl.gov}
\affiliation{Los Alamos National Laboratory, P.O. Box 1663, Los Alamos, NM 87545, U.S.A.}

\author{N. R. Shaffer}
\affiliation{Los Alamos National Laboratory, P.O. Box 1663, Los Alamos, NM 87545, U.S.A.}

\author{D. Saumon}
\affiliation{Los Alamos National Laboratory, P.O. Box 1663, Los Alamos, NM 87545, U.S.A.}

\author{R. Perriot}
\affiliation{Los Alamos National Laboratory, P.O. Box 1663, Los Alamos, NM 87545, U.S.A.}

\author{T. Nelson}
\affiliation{Los Alamos National Laboratory, P.O. Box 1663, Los Alamos, NM 87545, U.S.A.}

\author{L. A. Collins}
\affiliation{Los Alamos National Laboratory, P.O. Box 1663, Los Alamos, NM 87545, U.S.A.}

\author{C. Ticknor}
\affiliation{Los Alamos National Laboratory, P.O. Box 1663, Los Alamos, NM 87545, U.S.A.}

\date{\today}
\begin{abstract}
A new model for the electrical conductivity of dense plasmas with a mixture
of ion species, containing no adjustable
parameters, is presented.  The model takes the temperature, mass density
and relative abundances of the species as input.  It takes into account
partial ionization, ionic structure, and core-valence orthogonality, and
uses quantum mechanical calculations of cross sections.  Comparison to an
existing high fidelity but computationally expensive method reveals good 
agreement.  The new model is computationally efficient and can reach 
high temperatures.  A new mixing rule is also presented that gives
reasonably accurate conductivities for high temperature plasma mixtures.
\end{abstract}
\pacs{ }
\keywords{electrical conductivity, mixtures, dense plasmas}
\maketitle

\section{Introduction}
The calculation of the electrical conductivity of mixtures in the
dense plasma regime is a difficult challenge that only a handful of
studies have addressed.  The issue is complicated by
having to accurately model partial ionization, ionic structure, core-valence
orthogonality, and a wide range of coupling strengths (between the ions and other ions,
as well as between electrons and ions).  The electrical conductivity of such plasmas, and the
closely related thermal conductivity, are relevant
to inertial and magneto-inertial fusion experiments \cite{gomez14, rinderknecht18},
as well as to white dwarf stars \cite{marshak40}.

The ab initio method known as density functional theory molecular dynamics (DFT-MD),
in conjunction with the Kubo-Greenwood formula 
\cite{kohn,hohenberg,mermin65,greenwood58,desjarlais02, hu15}, can be used to
calculate the conductivity of mixtures for temperatures
near or below the Fermi energy.  This method uses few approximations but quickly becomes 
prohibitively expensive as temperature increases \cite{sjostrom15}.  

More approximate methods include the use of mixing rules to generate a mixture
conductivity from pure (single ion species) plasma conductivities \cite{starrett12a, johnson, johnson2}, but such an
approach necessarily ignores cross-species correlations.
Another approach is to treat the mixture as an effective single species \cite{faussurier14, more88}.
The early work of reference \cite{perrot95} used an extended Ziman formula
with local pseudopotentials to calculate the conductivity of mixtures in
the Born approximation with the Ziman approximation.  The Ziman method is suited 
to highly degenerate electronic systems, which is not generally the case for
plasmas.  The Born approximation can lead to large errors in the conductivity 
for strong scatterers \cite{burrill16}.  Other methods,
such as the Zubarev generalized linear response method could be used
to study the conductivity of mixtures \cite{zubarev73,redmer99}.

In this work we extend the previously developed potential of mean force method
\cite{starrett17} to plasma mixtures.  The extension introduces no additional
approximations.  We calculate the potential of mean force using the model
of reference \cite{starrett14b}, which is a DFT based average atom model that
couples to the quantum Ornstein-Zernike equations.  The cross sections
are calculated using the quantum mechanical t-matrix method.  We use the relaxation
time approximation \cite{bhatnager54,krall73} to calculate the conductivity, though
the cross sections could be used in other kinetic theory approaches such as the Ziman method 
\cite{ziman60}, or the Zubarev method \cite{zubarev73}.  We note that 
the use of the potential of mean force for transport was first developed
in reference \cite{baalrud13}, for classical ionic transport coefficients.

Results from the model are compared to DFT-MD results for a range of mixtures, in particular
to those relevant to inertial confinement fusion.  The results are in generally good agreement
and the model appears to be as accurate as the single species plasma model \cite{starrett17}.
The model is numerically convenient, taking a few minutes on a single processor per
point, and therefore is suitable for making data tables.
A new conductivity mixing rule is also presented and compared to the model.  The mixing rule 
takes the conductivity and average ionization of pure plasmas of each mixture species, at 
the same mass density and temperature, as input.  Good agreement between the mixing rule
and the potential for mean force model is found for high temperature plasmas.  

The structure of this paper is as follows.  In section \ref{sec_vmf} we review
the quantum Ornstein-Zernike equations for mixtures, and introduce the
potential of mean force, deriving it for mixtures.  We also transform
it into a numerically convenient and physically transparent form.
In section \ref{sec_relax} we give the equations to solve the relaxation
time approximation for mixtures.  In section \ref{sec_nr} numerical results
are presented.  These include comparisons to DFT-MD simulations for DT, BeDT and
C$_7$H$_9$.  The conductivity of asymmetric mixtures (high-Z with low-Z) is
investigated and the accuracy of a Thomas-Fermi potential of mean force model
is compared to the Kohn-Sham version.  Finally in section \ref{sec_con} we present our conclusions.
Throughout this work we use Hartree atomic units unless otherwise stated, in which
$\hbar = m_e = k_B = e = 1$, where the symbols have their usual meaning.

\section{Potential of mean force for mixtures\label{sec_vmf}}

\subsection{The potential of mean force}
The potential of mean force $V_{ij}^{MF}({\br })$ is the effective potential felt by
a non-interacting particles such that the resulting particle density $n_j({\br })$ is the same as in
the interacting system \cite{percus62}.  
For example, with classical particles the density distribution of species $j$ around a central particle 
of species $i$ can be written in the form \cite{hansen1}
\begin{flalign}
n_{j}(\br)  & =  n_j^0 \exp\left(  -\beta V_{ij}^{MF}(\br) \right)
\end{flalign}
and for quantal particles the expression is
\begin{flalign}
n_{e}(\br)  & =  \sum\limits_j f(\epsilon_j) |\phi_j(\br)|^2
\end{flalign}
where $f(\epsilon_j)$ is the Fermi-Dirac function,  $\phi_j(\br)$ is the wave function
that satisfies the Schr\"odinger equation with potential $V_{ie}^{MF}(\br)$, and
the sum is over all eigenstates.

To derive an expression for $V^{MF}_{ij}(r)$, we note that
if $F^{ex}$ is the excess free energy of the system, then
the Euler equations \cite{anta00} give
\begin{flalign}
V_{ij}^{MF}(\br) & = V_{ij}(r) + \frac{\delta F^{ex}}{\delta n_i(\br)}  -  \mu^{ex}_i 
\end{flalign}
where $\mu_i^{ex}$ is the excess chemical potential of species $i$
and $V_{ij}(r)$ is the pair interaction potential. 
By using a functional Taylor expansion of the excess free energy $F^{ex}$ 
about a uniform reference system, an expression for the potential of mean force $V_{ij}^{MF}(r)$ 
of an isotropic system can be found \cite{anta00,louis02}
\begin{eqnarray}
\!\!\!\!\!\! V^{MF}_{ij}(r)& \!\!\!=  \!\!\!& V_{ij}(r) + \sum\limits_{\lambda=1}^{N+1}
               n_\lambda^0 \int d^3 r^\prime 
               \frac{ C_{\lambda j}(|\br -\brp|) }{-\beta} h_{i \lambda }(r^\prime) 
\label{vmfij}               
\end{eqnarray}
where the direct correlation function is related to the second functional
derivative of the excess free energy
\begin{flalign}
C_{ij}(|\br -\brp|)
& = - \beta  
\frac{\delta^2 F^{ex}}{\delta n_i(\br) \delta n_j(\brp)}
\end{flalign}
For classical
ions of charge $\bar{Z}_i$ and $\bar{Z}_j$, $V_{ij}(r) = \bar{Z}_i \bar{Z}_j / r$.
For quantal electrons interacting with an ion of species $i$, 
$V_{ie}(r) = -\bar{Z}_i / r$ if we assume a point ion.  However, 
we define an ion as a point nucleus with bound electrons,
\begin{equation}
V_{ie}(r) = -\frac{Z_i}{r} + \int d^3r^\prime \frac{n_{i,e}^{ion}(r^\prime)}{\left| \br - \brp \right|}
+ V^{xc}[{n_{i,e}^{ion}}(r)]
\end{equation}
where $n_{i,e}^{ion}(r)$ is the electron density of the bound electrons around ion
$i$ and $V^{xc}$ is the exchange and correlation potential.

Implicit in equation (\ref{vmfij}) is the neglect of expansion terms beyond
second order.  For classical particles these higher order terms are
often collected together in the so-called bridge function \cite{hansen1, iyetomi82}.
Neglect of these terms corresponds the Hyper-Netted Chain (HNC) approximation \cite{morita58}.
%For quantal particles Chihara  named the neglect of these terms 
%the quantum Hyper-Netted Chain (QHNC) approximation \cite{chihara91}.  This 
%should not be confused with the QHNC model developed by Chihara \cite{chihara91}.
\subsection{Quantum Ornstein-Zernike equations}
The quantum Ornstein-Zernike (QOZ) equations \cite{chihara84a} for a mixture of quantal electrons and $N$
classical ion species were derived in reference \cite{starrett14b}
\begin{eqnarray}
\!\!h_{ij}(k)\! &\!\! =\!\! &\!\! \left(  -\frac{\chi_{jj}^{0}(k)}{\beta n_j^0} \right) 
\left[ C_{ij}(k) + \sum\limits_{\lambda=1}^{N+1} n_\lambda^0 h_{i\lambda}(k) C_{\lambda j}(k)\right]
\label{qoz_general}
\end{eqnarray}
where $h_{ij}(k)$ is the total correlation function in Fourier space ($h_{ij}(r) = g_{ij}(r) - 1$,
where $g_{ij}(r)$ is the pair distribution function), $C_{ij}(k)$ is the direct correlation function,
and $\chi_{jj}^0(k)$ is the non-interacting response function.  For classical particles 
$\chi_{jj}^0 / \beta n_j^0 = 1$, where $n_j^0$ is the average particle density for 
species $j$  and $\beta$ is the inverse of the temperature.  For the quantal 
electrons $\chi_{ee}^0$ is the finite-temperature Lindhard function \cite{lindhard54}.
Equation (\ref{qoz_general}) is valid if at least one of the species is classical.  For the
electron-electron pair distribution function, see reference \cite{shaffer19}.

To solve the QOZ equations we map them to a system of $N$ classical ions screened by electrons, 
where the screening density for species $i$ is
\begin{eqnarray}
n_{i,e}^{scr}(k) = -\frac{C_{i e}(k)}{\beta} \chi_{ee}^\prime(k)
\label{scr_den_def}
\end{eqnarray}
with
\begin{equation}
\chi_{ee}^\prime(k) \equiv
\frac{\chi_{ee}^{0}(k) }
{ 1 + \chi_{ee}^{0}(k) C_{ee}(k) / \beta}.
\label{chip}
\end{equation}
This mapping leads to the $N$-component classical OZ equations with HNC closure relations \cite{percus62}
\begin{flalign}
h_{IJ}(k) & =  C_{IJ}(k) + \sum\limits_{\lambda=1}^{N} n_\lambda^0 h_{I\lambda}(k) C_{\lambda J}(k) \\
h_{IJ}(r) + 1 & =  \exp\left(  -\beta V_{IJ}(r) + h_{IJ}(r) - C_{IJ}(r) \right) \label{c_hij}
\end{flalign}
where the ion-ion pair interaction potentials are given by
\begin{eqnarray}
V_{I J}(k) & = & \frac{ 4 \pi \bar{Z}_i \bar{Z}_j }{k^2} - \frac{C_{i e}(k)}{\beta} n_{j,e}^{scr}(k).
\label{coz_pot}
\end{eqnarray}
Here, screened ion $I$ ($J$) corresponds to ion $i$ ($j$) in the full QOZ equations.
These pair potentials, equation (\ref{coz_pot}), are the interaction potentials for pairs of
`dressed' ions, defined by a point charge $\bar{Z}_j$ and a 
neutralizing, screening electron cloud $n_{j,e}^{scr}(r)$.
In the limit of high temperatures equation (\ref{coz_pot}) reduces to a Deybe-H\"uckel potential
with the screening length given by the electron Dybye length.

%Equation (\ref{c_hij}) is found using equations (\ref{}), (\ref{}), and (\ref{})
The screening densities are provided by finite temperature density functional theory \cite{mermin65}
average atom calculations \cite{starrett14b}.  One can view these screening densities as electron-ion
closure relations.  
The closure relation for the electron-electron direct correlation function $C_{ee}$ is provided
by the jellium model, with the local field correction provided by reference \cite{chabrier90}.
With these closure relations we can solve the QOZ's for the correlation functions $h_{ij}$
and $C_{ij}$.
%It is worth noting that in Chihara's Quantum Hyper-Netted Chain approximation 
%\cite{chihara91} the electron-ion closure relation is provided by using and average atom to provide
%$g_{ie}(r)$ directly.  We found this to be more numerically challenging.  Moreover, Chihara's
%closure model requires the use of a so-called ``chemical'' model and this is difficult to 
%generalize in the plasma regime.  Our approach uses a so-called ``physical'' model \cite{starrett14b}.

\subsection{Electron-ion potential of mean force}
Equation (\ref{vmfij}) is valid as long as at least one species is classical \cite{louis02,shaffer19}. The electron-ion potential of mean
force is then
\begin{eqnarray}
V^{MF}_{ie}(r)& =  & V_{ie}(r) + \sum\limits_{\lambda=1}^{N}
               n_\lambda^0 \int d^3 r^\prime 
               \frac{ C_{\lambda e}(|\br -\brp|) }{-\beta} h_{i \lambda }(r^\prime) \nonumber\\
&& +
               \bar{n}_e^0 \int d^3 r^\prime 
               \frac{ C_{ee}(|\br -\brp|) }{-\beta} h_{i e}(r^\prime) 
\end{eqnarray}
\begin{figure}
\begin{center}
\includegraphics[scale=0.35]{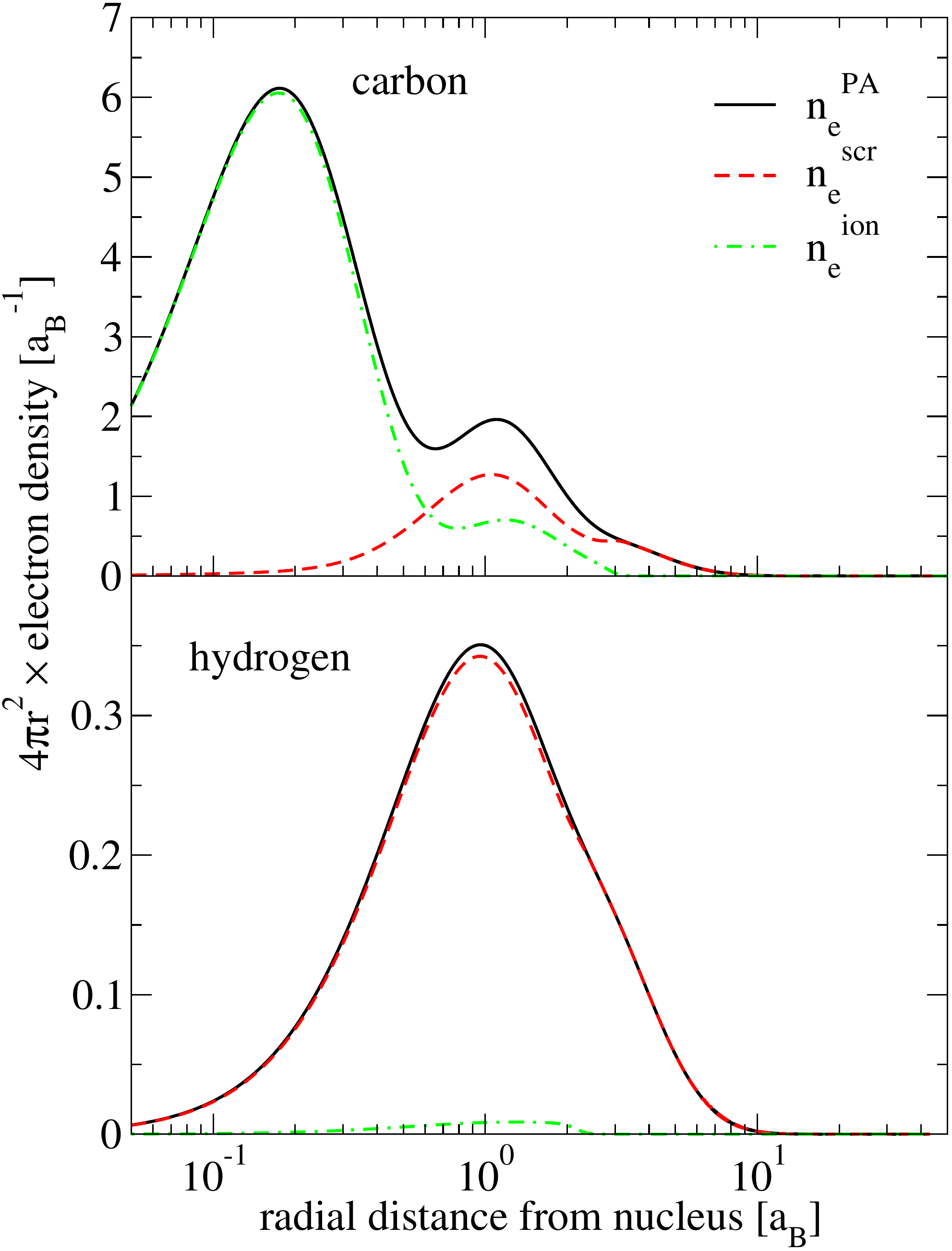}
\end{center}
\caption{(Color online) Example of pseudoatom, screening and ion densities for
an equi-molar carbon-hydrogen mixture at 1 g/cm$^3$ and 10 eV temperature. For
carbon, average ionization per atom is $\bar{Z}_C=3.01$, for hydrogen it is 
$\bar{Z}_H=0.986$. 
}
\label{fig_pa}
\end{figure}
\begin{figure}
\begin{center}
\includegraphics[scale=0.35]{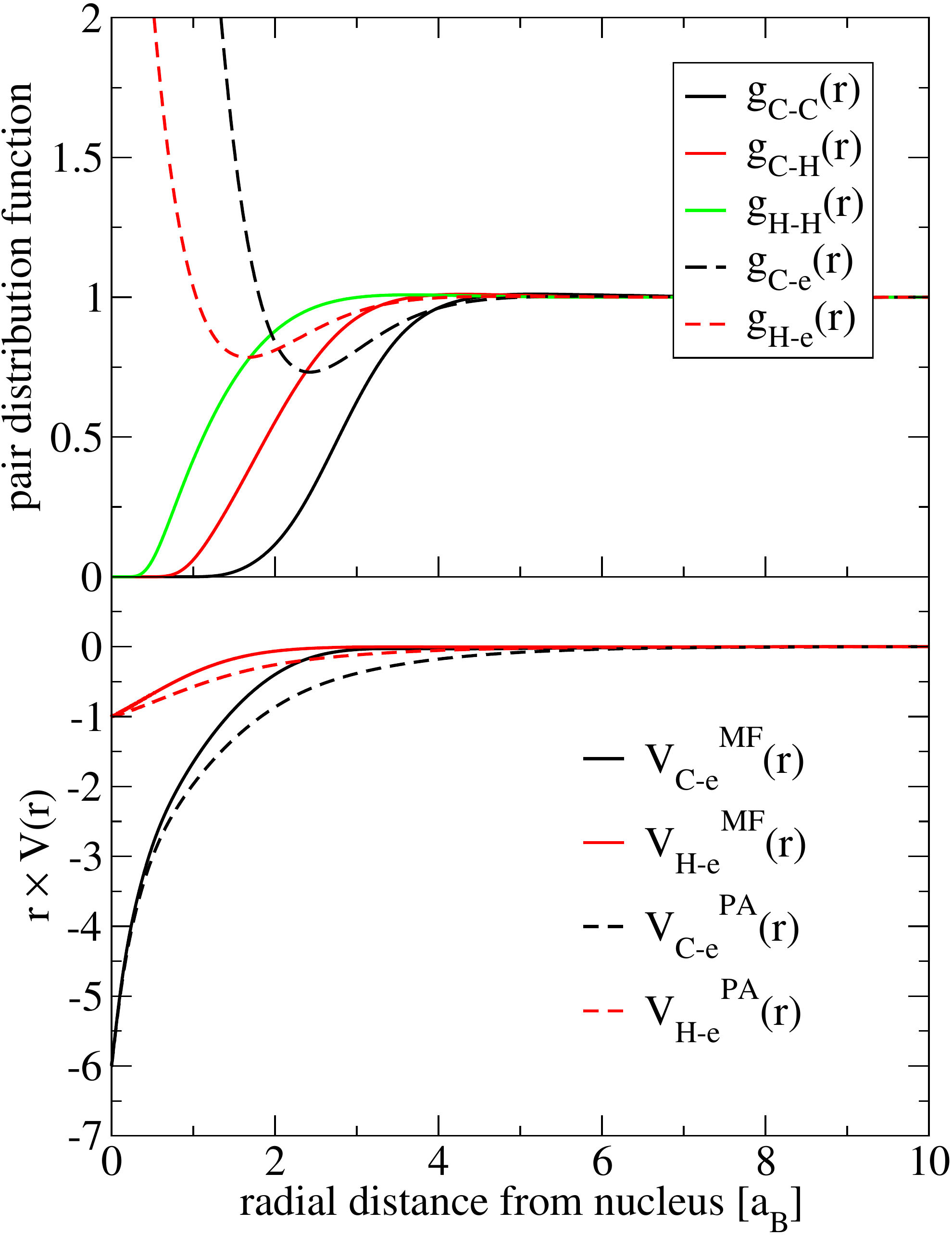}
\end{center}
\caption{(Color online) Example of ion-ion and ion-electron pair distribution
functions (top panel) for an equi-molar carbon-hydrogen mixture at 1 g/cm$^3$ 
and 10 eV temperature.  The bottom panel shows the potentials $V^{MF}(r)$
and $V^{PA}(r)$ for the same case.  The difference is the contribution to
screening from the other ions and their screening electrons.
}
\label{fig_vmf}
\end{figure}
Setting $N=1$ recovers the single species result, equation (26) of reference \cite{starrett17}.
This potential can be rewritten as a sum of Hartree, ion-ion and ion-electron correlation, and
electron-electron exchange and correlation terms.  First note that the QOZ equations give
\begin{flalign}
\bar{n}_e^0 h_{ie}(k) 
 = & n_{i,e}^{scr}(k) + \sum\limits_{\lambda=1}^{N} n_\lambda^0 h_{i\lambda}(k) n_{\lambda, e}^{scr}(k)
\label{dnie}
\end{flalign}
Now define
\begin{eqnarray}
n^{x}_{i,e}(k) & = & \sum\limits_{\lambda=1}^{N} n_\lambda^0 h_{i\lambda}(k) n_{\lambda, e}^{scr}(k) 
%n^{x}_{i,e}(r) & = & \sum\limits_{\lambda=1}^{N} n_\lambda^0 \int d^3 r^\prime h_{i\lambda}(|\br-\brp|) n_{\lambda, e}^{scr}(r^\prime) 
\end{eqnarray}
which is the denisty of electrons surrounding the non-central ions, then
\begin{flalign}
\bar{n}_e^0 h_{ie}(k) 
 = & n_{i,e}^{scr}(k) + n^{x}_{i,e}(k) 
\label{dnie2}
\end{flalign}
Using this together with
\begin{equation}
C_{ij}(k) =  -\beta V_{ij}^C(k) + \widetilde{C}_{ij}(k) 
\end{equation}
where $\widetilde{C}_{ij}(k)$ contains the correlation contribution (and exchange for $\widetilde{C}_{ee}$),
and $V_{ij}^C$ is the Coulomb pair potential, then
\begin{eqnarray}
V_{ie}^{MF}(r) & = & V_i^{PA}(r) +  \int d^3 r^\prime 
\frac{ \sum\limits_{\lambda =1}^{N} - n_\lambda^0 \bar{Z_\lambda} h_{\lambda i}(r^\prime) + n_{i,e}^{x}(r^\prime)}{|\br -\brp|} \nonumber\\
          &   & + V^{xc}[n_{i,e}^{x}(r)+\bar{n}_e^0] - V^{xc}[\bar{n}_e^0]\nonumber\\
          &   & + \sum\limits_{\lambda=1}^N n_\lambda^0 \int d^3 r^\prime 
                \frac{ \widetilde{C}_{\lambda e}(|\br -\brp|) }{-\beta} h_{i\lambda}(r^\prime)
          \label{vmf}
\end{eqnarray}
Here
\begin{equation}
V_i^{PA}(r) = -\frac{Z_i}{r} + \int d^3r^\prime \frac{n_{i,e}^{PA}(r^\prime)}{\left| \br - \brp \right|}
+ V^{xc}[n_{i,e}^{PA}(r)]
\end{equation}
is the pseudoatom potential and
\begin{equation}
n_{i,e}^{PA}(r) = n_{i,e}^{ion}(r) + n_{i,e}^{scr}(r).
\end{equation}
is the pseudoatom density, which here is provided by the average atom
model \cite{starrett14b}.  The pseudoatom is defined by the nucleus and the electrons
that screen it, including bound and conduction electrons.  
The second term in equation (\ref{vmf}) is Hartree-Coulomb interactions between the ions
surrounding the central pseudoatom, and the electrons surrounding these ions ($n_{i,e}^x$).
The third and fourth terms are the exchange and correlation interactions of these electrons.
The fifth term is due to electron-ion correlations.

An example of pseudoatom, 
screening (conduction) and ion (bound) densities are shown in figure \ref{fig_pa}.  For the same case, the
pair distribution functions and potentials of mean force are shown in figure
\ref{fig_vmf}.
The peaks in the carbon pseudoatom density in figure \ref{fig_pa} are due to
the shell structure;  the peak near 0.2 a$_B$ is due to electrons in the $1s$
orbital, the peak near 1 a$_B$ is due to electrons in the $2s$ orbital.  The
tail of the electron density is due to the ionized electrons.
For hydrogen, the $1s$ state is weakly bound $\epsilon_{1s} = -0.0042$ E$_h$ and
is nearly fully depopulated, leading to a small bound electron density around each hydrogen nucleus 
$n_{i,e}^{ion}(r)$ (figure \ref{fig_pa}).
The ionized electrons form the screening density which dominates the pseudoatom
density.

In the top panel of figure \ref{fig_vmf}, the ion-electron pair distribution functions
for carbon and hydrogen reflect the smaller size and charge of the hydrogen ion.  This feeds
into the ion-ion pair distribution functions, where the Coulomb hole, due to
repulsion of the ions to each other at small separations,  is smaller for
H-H than for C-C distributions, with C-H lying in between.  The lack of oscillations
in the ion-ion pair distribution functions indicates a moderately coupled
ionic fluid, as expected under these conditions.  The bottom panel of
the figure shows the ion-electron potentials of mean force compared
to the pseudoatom potentials.  The effect of including the screening
from the other ions is to weaken the effective scattering potential, i.e. make
the effective screening length shorter.  

\section{electrical conductivity with the relaxation time approximation \label{sec_relax}}
\begin{figure}
\begin{center}
\includegraphics[scale=0.3]{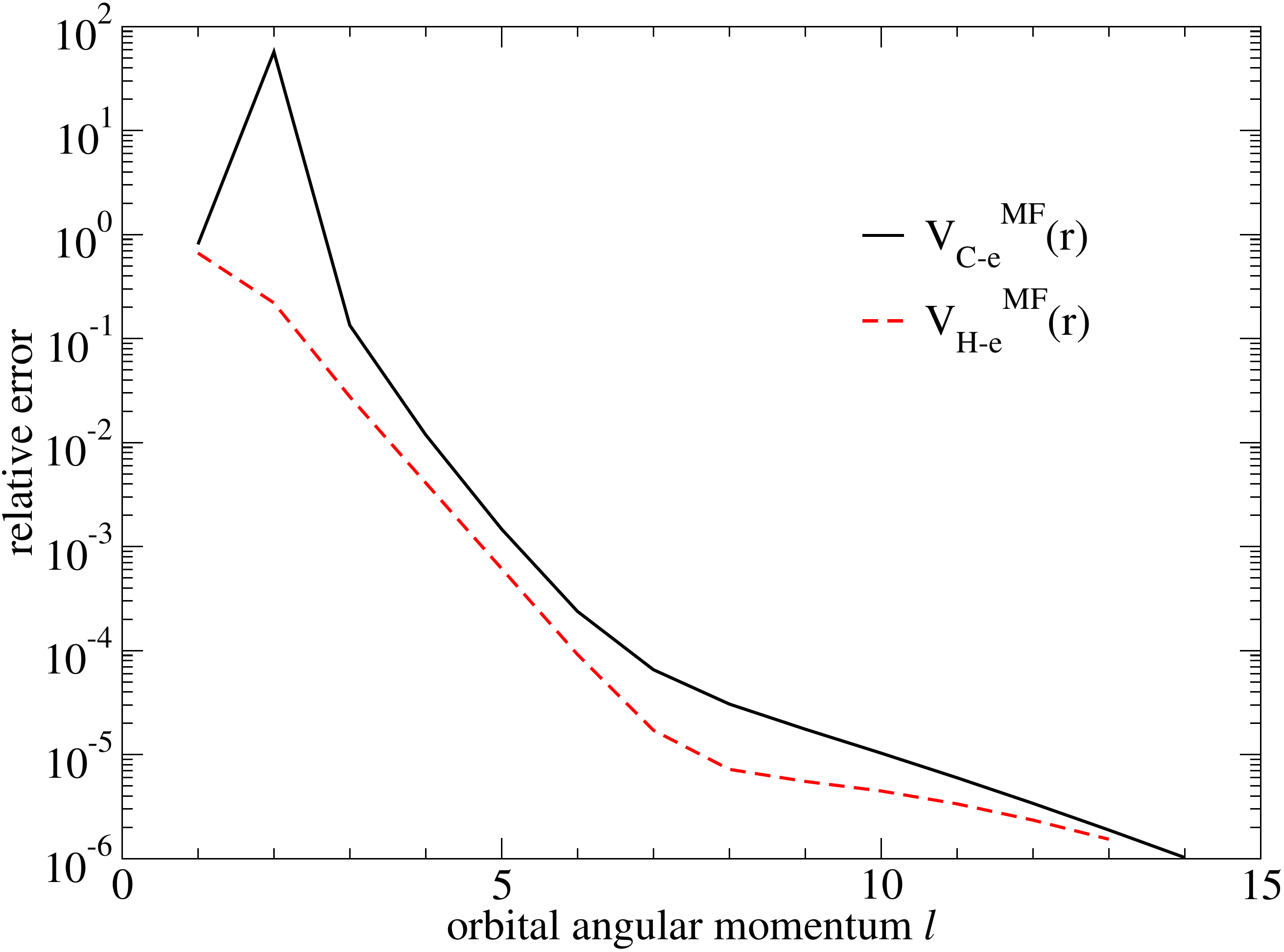}
\end{center}
\caption{(Color online) Convergence of the conductivity with respect to the maximum
orbital angular momentum for the sum in equation (\ref{trq}), for 
an equi-molar carbon-hydrogen mixture at 1 g/cm$^3$ and 10 eV temperature.
We converge on the relative error in the effective conductivity for each species, hence
the two lines.  More degenerate plasmas should converge faster, while more weakly degenerate
cases will converge more slowly, for $l>30$ we use a semi-classical calculation of
the phase shifts \cite{starrett17} for computational efficiency.
}
\label{fig_l}
\end{figure}

We calculate the electrical conductivity of the plasma using the relaxation time
approximation \cite{bhatnager54,krall73}.
\begin{equation}
\sigma_{DC} = \frac{1}{3\pi^2} \int_0^\infty \left( -\frac{df(\epsilon, \mu)}{d\epsilon} \right)  v^3 \tau_\epsilon d\epsilon
\label{sdc}
\end{equation}
where $f(\epsilon,\mu)$ is the Fermi-Dirac occupation factor, $\epsilon$ the electron energy $\epsilon = m v^2 /2$, 
$\mu$ is the electron chemical potential and $\tau_\epsilon$ is the relaxation time.
Using Matthiessen's rule, which assumes that scattering mechanisms are independent \cite{kasap17}, 
$\tau_\epsilon$ is calculated from the relaxation time due to each species \cite{lee84}
\begin{equation}
\frac{1}{\tau_\epsilon} = \sum\limits_{i=1}^{N} \frac{1}{\tau_{i,\epsilon}}
\label{rt}
\end{equation}
where the relaxation times are calculated from the momentum transport cross sections
\begin{equation}
\tau_{i,\epsilon} = \frac{1}{n_i^0\,v\, \sigma_{i,\mathrm{tr}}(\epsilon) }
\end{equation}
The momentum transport cross section is calculated by solving the Schr\"odinger equation
for the phase shifts $\eta_l(\epsilon)$ due to the scattering potential $V_{ie}^{MF}(r)$
\begin{equation}
\sigma_{i,\mathrm{tr}}(\epsilon) = \frac{4 \pi}{v^2} \sum\limits_{l=0}^{\infty} (l+1) \left( \sin\left( \eta_{l+1} - \eta_{l} \right)\right)^2
\label{trq}
\end{equation}
where the sum over orbital angular momentum quantum number $l$ converges, see figure \ref{fig_l}.
In the figure we show the relative change in the effective one-species conductivity caused by
adding another term to the $l$ summation in equation (\ref{trq}).
The effect of electron-electron collisions is modeled using the fit formula
of reference \cite{reinholz15}.

We note that equations (\ref{sdc}) and (\ref{rt}) cannot be decomposed (without approximation) into a simple sum
of single species conductivities, as one would have in a mixing rule, for example \cite{starrett12a}.

\section{numerical results\label{sec_nr}}
\subsection{Comparison to DFT-MD simulations}
\begin{figure}
\begin{center}
\includegraphics[scale=0.35]{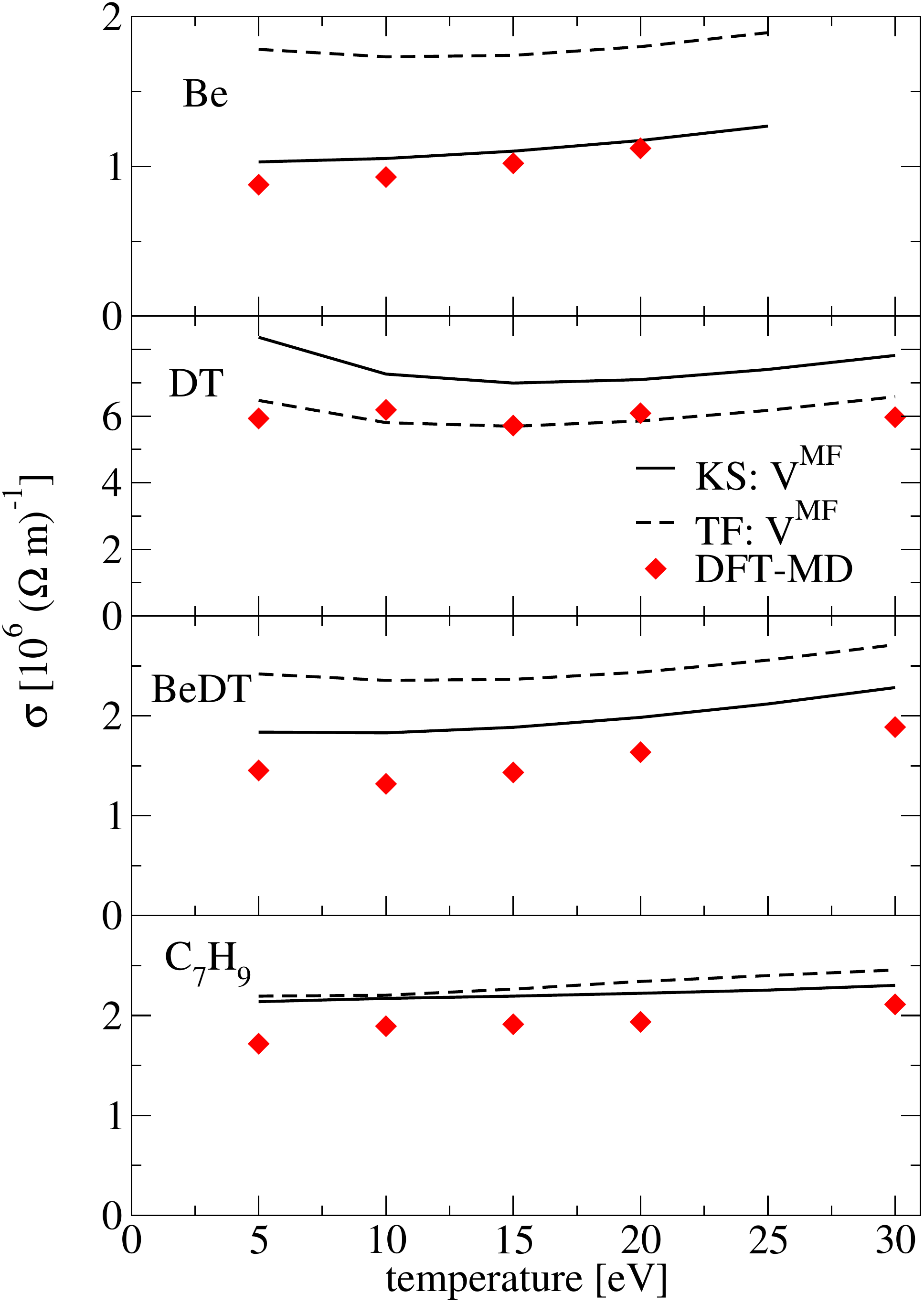}
\end{center}
\caption{(Color online) Results from the present model $V^{MF}$ in the Thomas-Fermi (TF)
or Kohn-sham (KS) approximation, compared to DFT-MD simulations using the Kubo-Greenwood 
approximation \cite{hanson11,starrett12a}.  In all cases the density is 10 g/cm$^3$.
The Fermi temperature for Be is 44.3 eV, for DT 65.0 eV, for BeDT 51.7 eV, and for CH 65.0 eV.
}
\label{fig_temp}
\end{figure}

In figure \ref{fig_temp} we compare results from the present model to DFT-MD
simulations that used the Kubo-Greenwood approximation to evaluate the
transport properties \cite{hanson11,starrett12a}.  Our calculations use either
the potential of mean force calculated using the semi-classical, orbital free, Thomas-Fermi (TF) version of
the model presented in reference \cite{starrett14b}, or the less approximate Kohn-Sham version (KS) .  In all
cases we have used the temperature dependent exchange and correlation potential of reference \cite{ksdt}.
In the caption we also give the Fermi temperature $T_F$ for each case (using the KS ionization value).  This depends
very weakly on temperature (via the ionization) so we give the average value for the conditions plotted.  
For all cases the temperature is much less than $T_F$
indicating a degenerate plasma.
\begin{figure}
\begin{center}
\includegraphics[scale=0.3]{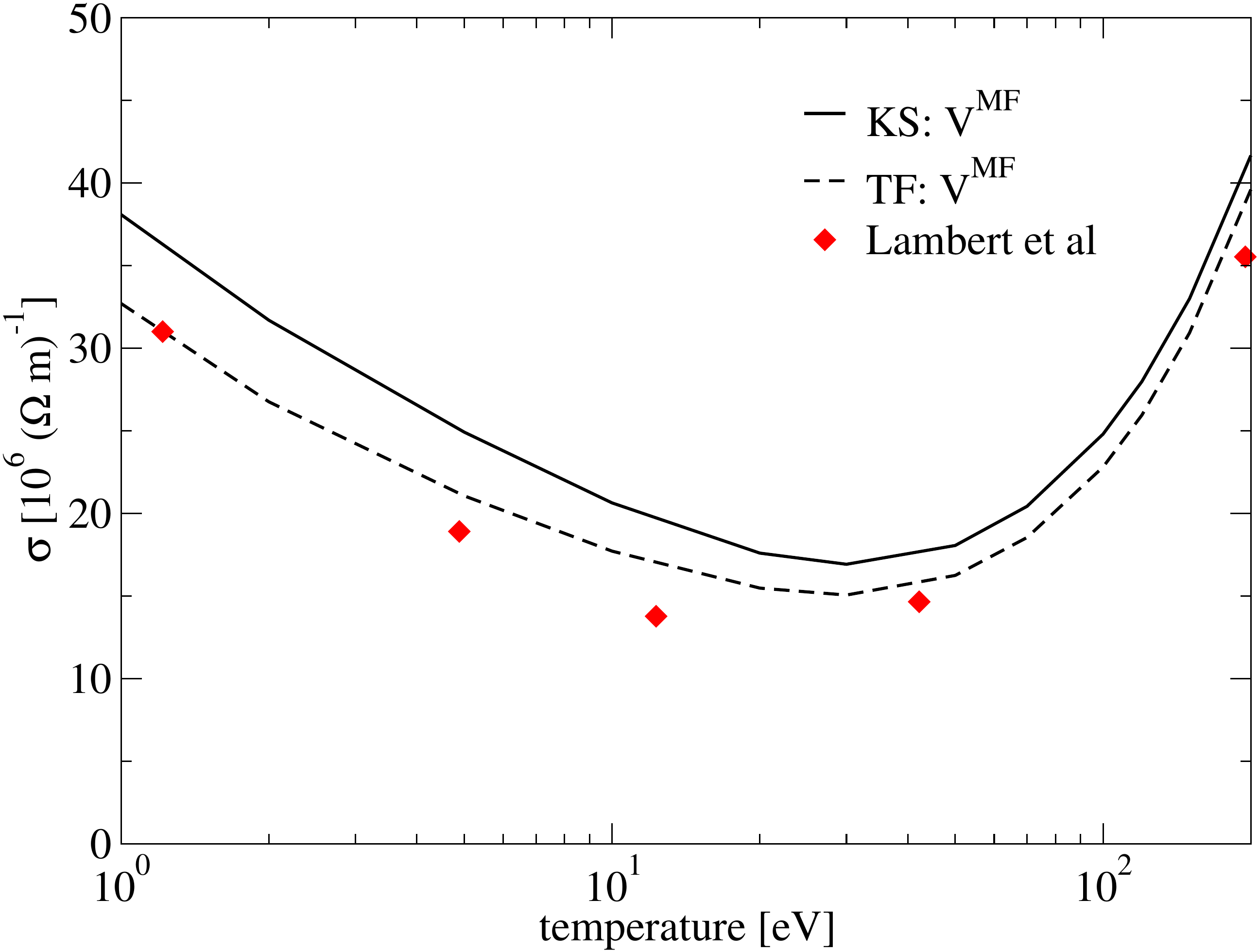}
\end{center}
\caption{(Color online) Results from the present model compared to the Kubo-Greenwood
DFT-MD results of reference \cite{lambert11} for dense hydrogen at 10 g/cm$^3$.
}
\label{fig_h}
\end{figure}

From figure \ref{fig_temp} we see reasonably good agreement of the present model with
the DFT-MD calculations.  In physical content these DFT-MD simulations are
more complete than the present model and therefore should be more accurate.  They are however,
expensive in terms of computational cost and can have issues with
numerical convergence \cite{lambert11, desjarlais17}.  This computational cost
limitation becomes increasingly acute for temperatures above the Fermi energy.  For
all cases except the DT mixture, the KS version of the model is in better agreement
with the DFT-MD than the TF version, as expected, as it is a less approximate method.
The exception, DT is further examined in figure \ref{fig_h}.  There we compare to
Kubo-Greenwood DFT-MD results for pure hydrogen at 10 g/cm$^3$ as a function of
temperature.  Again we see that the TF version of the model is in closer
agreement with the DFT-MD results.  We have no definite explanation of this,
but believe it just fortuitous.  The KS version overestimates the conductivity in
these cases, where $\bar{Z} = 1$ is predicted and is expected.  The error is reduced in the TF 
version because $\bar{Z} <  1$ due to the lack of shell structure.  Hence, a cancellation
of errors may be occurring for the TF version.  For Be, figure \ref{fig_temp}, TF overestimates the
conductivity and also predicts a larger average ionization $\sim 2.7$, versus $\sim 2.0$ for 
KS, so this is consistent.  We also note that good agreement of the present model 
with DFT-MD simulations for the electrical conductivity of hydrogen plasmas,
at 40 g/cm$^3$ and temperatures from 500 to 900 eV, was found in 
reference \cite{starrett17}.

\begin{figure}
\begin{center}
\includegraphics[scale=0.3]{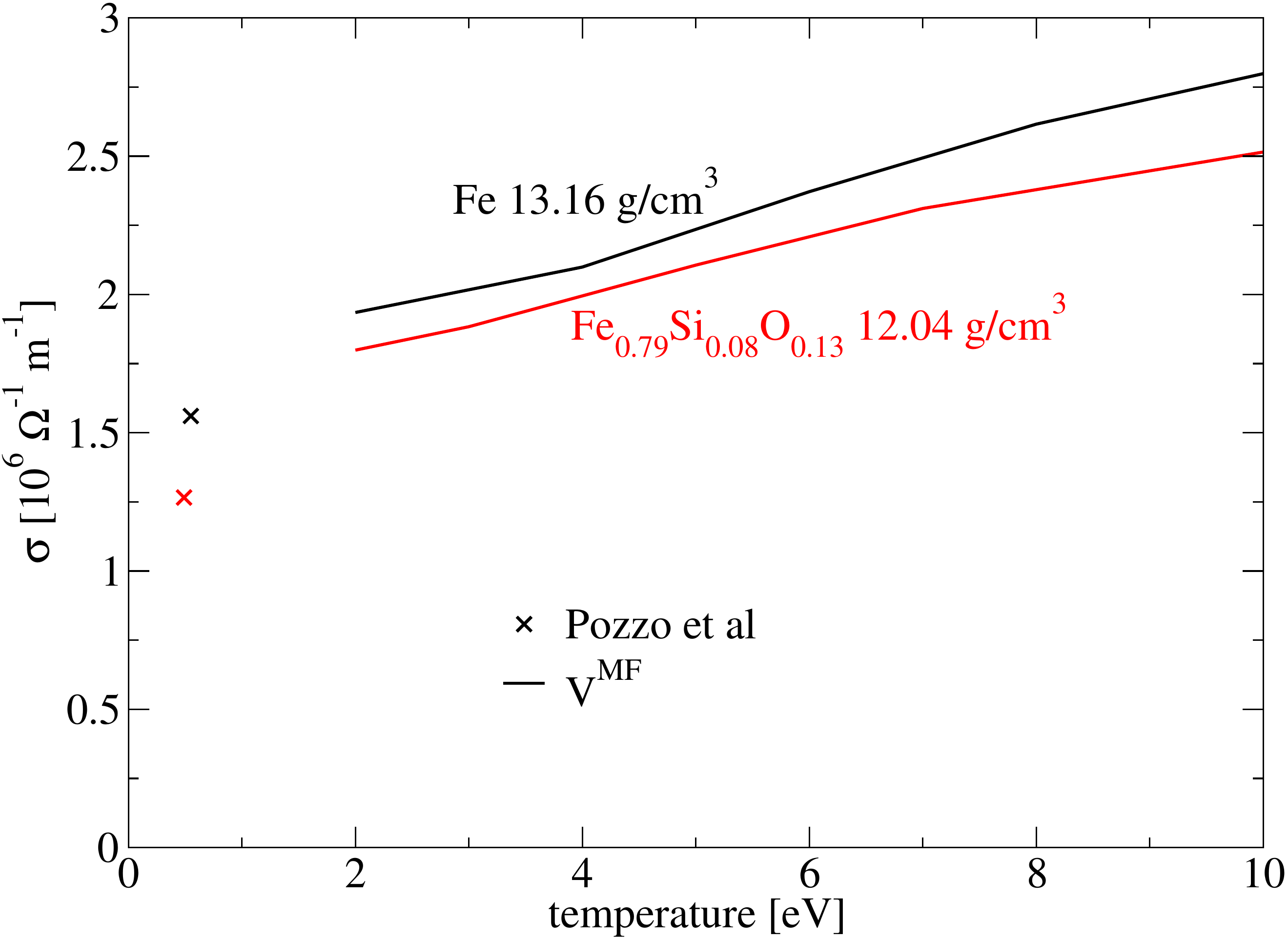}
\end{center}
\caption{(Color online) Comparison of the present model to the Kubo-Greenwood 
DFT-MD calculations of reference \cite{pozzo13}.  We use the KS version of our
model.
}
\label{fig_fe}
\end{figure}
In figure \ref{fig_fe} we compare to DFT-MD results for pure Fe and for a Fe mixture
relevant to the Earth's interior \cite{pozzo13}.  We were unable to obtain reliable results at 
the same temperature as the DFT-MD results.   
This is due to a breakdown of a number of approximations in the model.  On the one hand
the average atom itself is inaccurate under these conditions due to the
free electron boundary conditions on the atom, whereas for iron, multiple
scattering, which is ignored here, is strong \cite{starrett18}.   Also, to solve
the QOZ equations (and to use the kinetic theory model), a definition
of an ion is required.  This is difficult to do unambiguously due to the
large iron $3d$ resonance state in the continuum of free electrons.
Moreover, use of the relaxation time approximation (\ref{sdc}) assumes a binary collision
approximation, which becomes an unsafe assumption for resonance states,
as evidenced by the strong multiple scattering effect \cite{starrett18}.
All these effects conspire to mean the model is unreasonable for
dense iron at temperatures below $\sim 2$ eV. In contrast, for aluminum
\cite{gill19}, we found reasonable results to 0.2 eV at solid density,
where none of the above problems are relevant.
Nevertheless, the trend seen in figure \ref{fig_fe}, of a reduced conductivity 
for the mixture, is reproduced and the absolute numbers are reasonable.

\begin{figure}
\begin{center}
\includegraphics[scale=0.3]{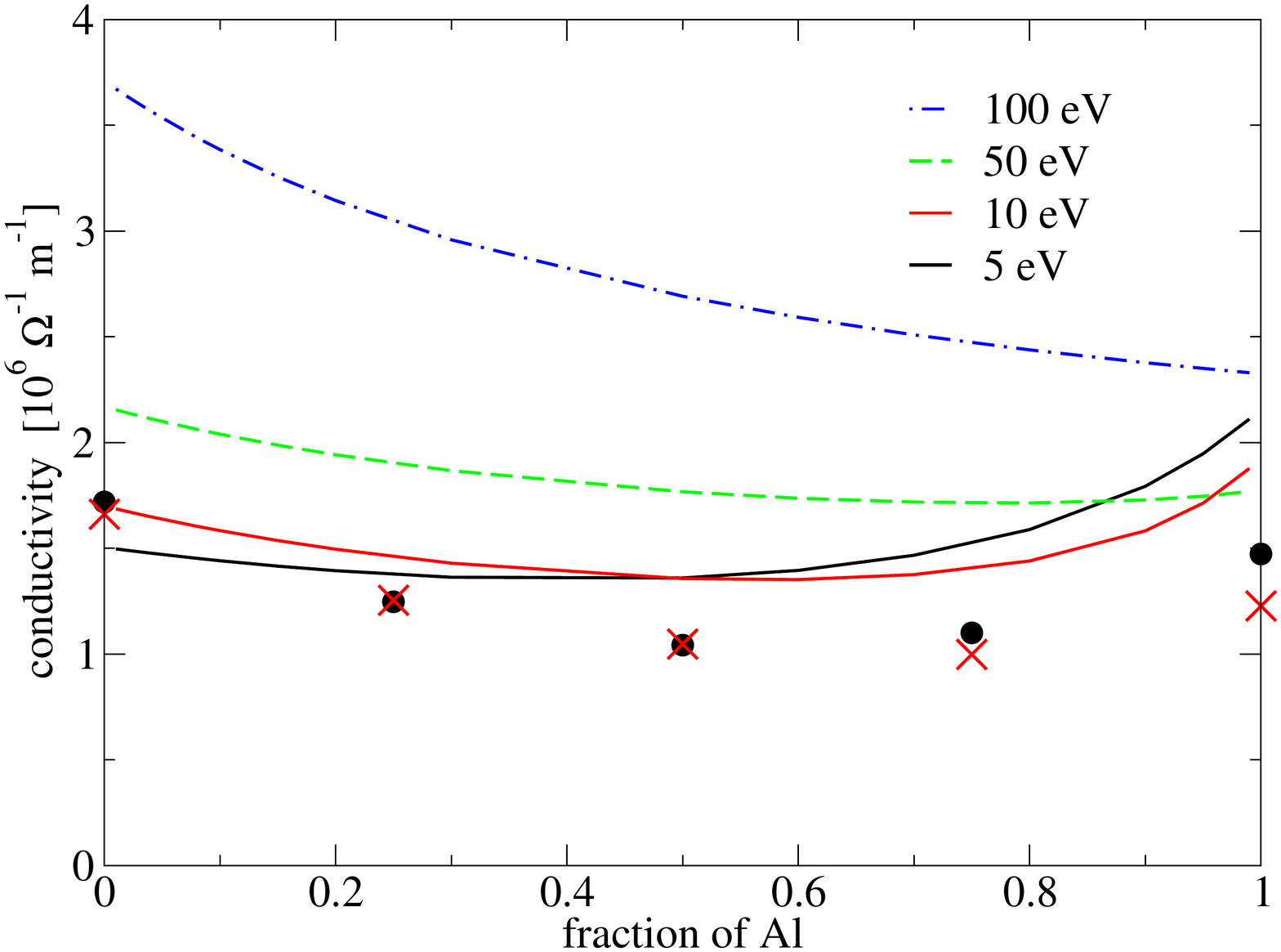}
\end{center}
\caption{(Color online) 
Conductivity of aluminum-carbon mixture at 10 g/cm$^3$.  Lines are from the
present model, symbols are Kubo-Greenwood DFT-MD results (filled circles 5 eV, crosses 10 eV).  
}
\label{fig_alc}
\end{figure}
In figure \ref{fig_alc} we compare the present model to Kubo-Greenwood DFT-MD
simulations for an aluminum-carbon mixture at 10 g/cm$^3$ (see appendix \ref{dft} for details of
our DFT-MD simulations).  
The level of agreement is reasonable at both 5 and 10 eV, 
and is similar to what we have seen in figures \ref{fig_temp} to \ref{fig_fe}.
Interestingly, for the two lowest temperatures a minimum in the conductivity 
is observed as a function of the fraction of aluminum in both the model and DFT-MD 
results.  This is reminiscent of
Nordheim's rule for alloys \cite{kasap17}.  The conductivity decreases
as the `impurity' is added starting for either pure phase.  Faber and Ziman
explained this behavior as being due to cross-terms in the scattering amplitudes \cite{faber65}.
These cross-terms are important if structure factors $S_{ij}(k)$ are significantly
different from unity at relevant electronic wave numbers $k$.  These relevant wave
numbers are determined by the range of values for which derivative of the Fermi occupation factor
is different from zero.  
In degenerate cases only the Fermi wave number $k_F$ is relevant, and the $S_{ij}(k_F)$ typically differ from 
unity.
At higher temperature,
larger $k$'s are relevant where $S_{ij}(k) \to 1$ so the conductivity changes
monotonically, as the cross-terms are negligible.
Physically, the minimum for more degenerate systems is related to a reduction in coherent scattering due to the
presence of impurities.   At the higher temperatures 
coherent scattering is disrupted by thermal ionic disorder.
For the two highest temperatures plotted in figure \ref{fig_alc}, the drop in conductivity 
on increasing aluminum fraction is mainly due to a decrease in the average ionization.
\begin{figure}
\begin{center}
\includegraphics[scale=0.3]{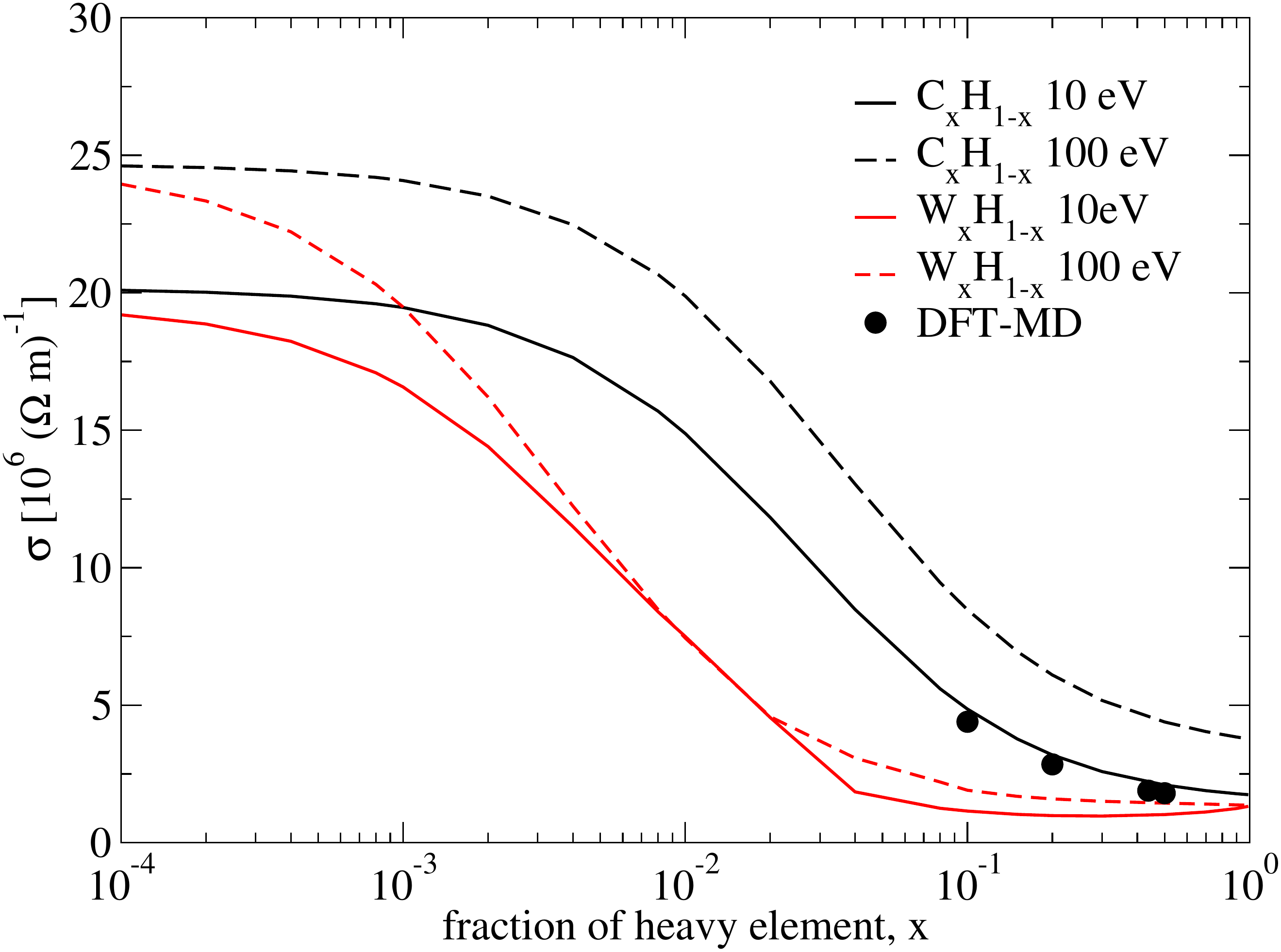}
\end{center}
\caption{(Color online) 
Electrical conductivity of high-Z, low-Z mixtures at 10 g/cm$^3$ as a function
of the fraction of the heavy element.  We have used the KS version of the model.
Also shown are DFT-MD Kubo-Greenwood calculations for C$_x$H$_{1-x}$ at 10 eV.
These are extensions of the calculations of reference \cite{starrett12a}.
}
\label{fig_ratio}
\end{figure}

Overall, the level of agreement seen in figures \ref{fig_temp} to \ref{fig_alc}
indicates that the mixture and the single species models are of similar accuracy, 
relative to Kubo-Greenwood DFT-MD.
We expect the predictions to become more accurate as temperature increases and
the model will eventually recover the Debye-H\"uckle limit \cite{ovechkin19}.

\subsection{Applications of the model}
Next we look at two applications of the model.  In figure \ref{fig_ratio}
we show the conductivity of C$_x$H$_{1-x}$ and W$_x$H$_{1-x}$ as a function
of $x$ for plasmas at 10 g/cm$^3$.  
Also shown are DFT-MD Kubo-Greenwood calculations for C$_x$H$_{1-x}$ at 10 eV. 
These are new calculations but are essentially extensions of the calculations
presented in reference \cite{starrett12a, hanson11}.  Good agreement
is observed.  We note that it becomes impractical to use DFT-MD for mixtures
where one element is a trace due to the need for at least one atom of the trace
species to be in the computational supercell, and preferably more than one atom
to reduce statistical noise.
The main result of figure \ref{fig_ratio} is a highly asymmetrical
transition between the single species plasmas, with the asymmetry being
more pronounced for the higher-Z mixture.
\begin{figure}
\begin{center}
\includegraphics[scale=0.35]{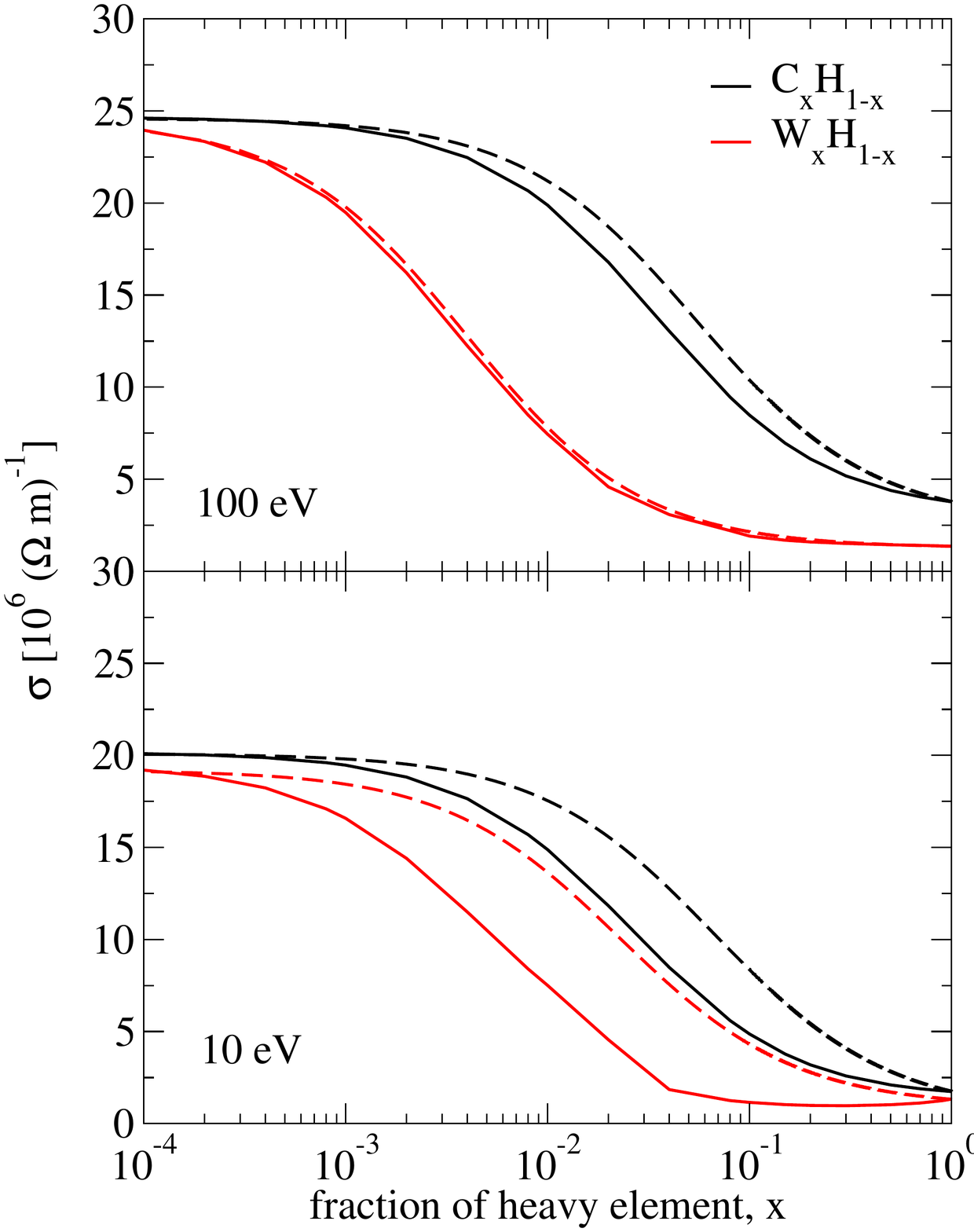}
\end{center}
\caption{(Color online) 
Solid lines:  the result from the KS version of the model, as in
figure \ref{fig_ratio}.  Dashed lines: the new mixing rule equation
(\ref{mr}).  Top panel, results at a temperature of 100 eV; bottom panel, results at 10 eV.  
The mixing rule works better at high temperature, as expected.
}
\label{fig_mix}
\end{figure}
\begin{figure}
\begin{center}
\includegraphics[scale=0.35]{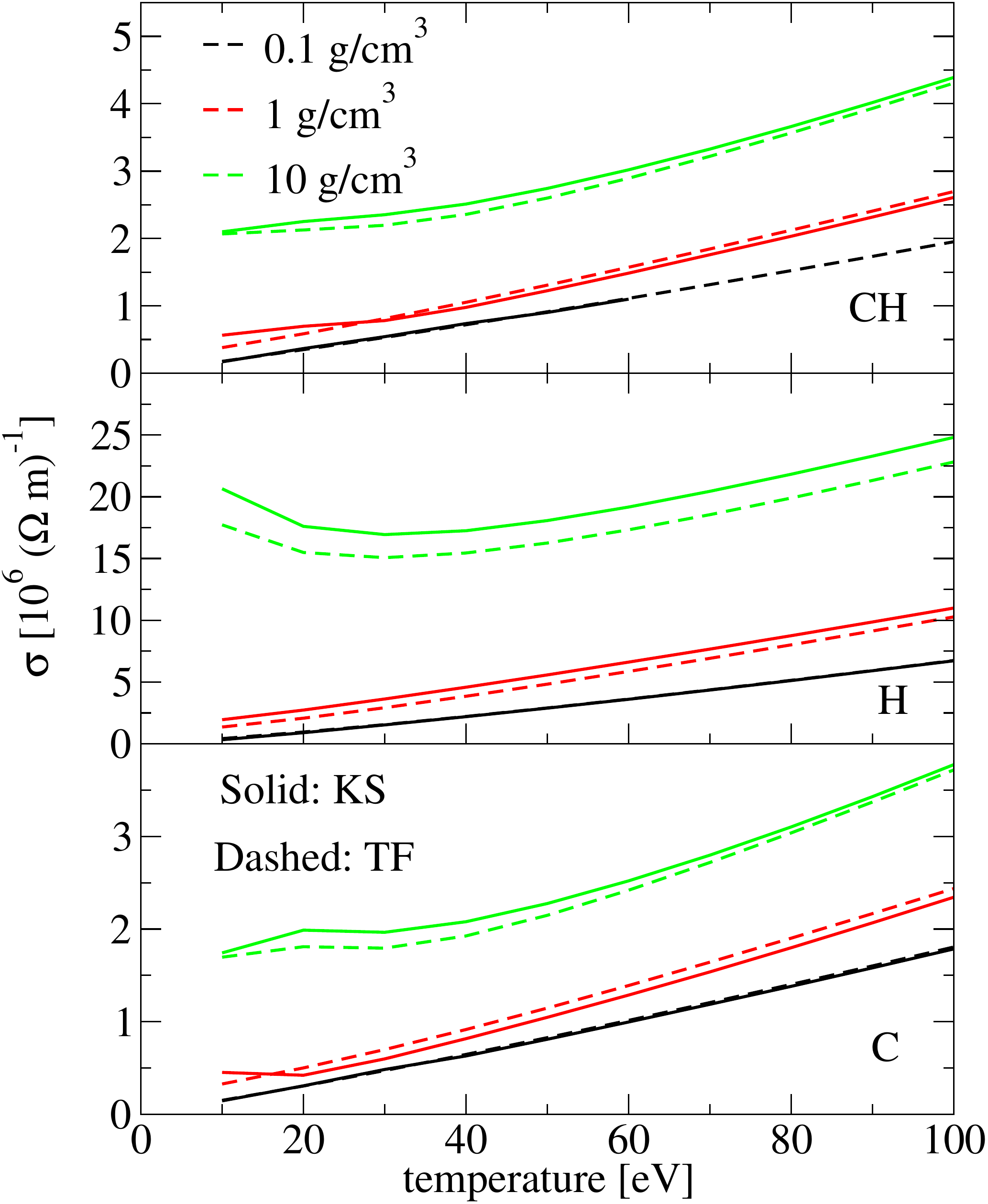}
\end{center}
\caption{(Color online) 
Comparison of the TF and KS versions of the model for CH, H and C
plasmas for 0.1, 1, and 10 g/cm$^3$, from 10 through 100 eV.
The TF and KS models agree best for non-degenerate cases.
}
\label{fig_compare}
\end{figure}
This behavior can be understood by considering the following
mixing rule (which is described in appendix \ref{mix})
\begin{equation}
\sigma_{DC} = \sum\limits_{i=1}^{N} Y_i \, \sigma_{i, DC} 
\label{mr}
\end{equation}
where $\sigma_{i, DC} $ is the conductivity of a pure (single ion species) plasma
of species $i$, and the coefficients $Y_i$ are
\begin{equation}
Y_i = x_i
\frac{{(\bar{Z}_i^p)}^2 }
{\sum\limits_{j=1}^{N} x_j {(\bar{Z}_j^p)}^2 }
\label{y2a}
\end{equation}
This mixing rule takes the conductivities ($\sigma_{i,DC}$) and average ionizations
$\bar{Z}_i^p$ of pure (indicated by superscript $p$) plasmas of the mixture components 
(at the same mass density and temperature), and generates the mixture conductivity $\sigma_{DC}$.

In figure \ref{fig_mix} we show the result of this new mixing rule for CH and WH plasmas.
We see that the trends are well reproduced telling us that the asymmetrical behaviour
can be understood as the weighted mixing of large and small
ion charges.  The mixing rule will become inaccurate for lower temperatures, and figure \ref{fig_mix} 
confirms this.  This is in part due to the breakdown of the assumed Coulomb logarithm form (appendix \ref{mix}).
The model (equation (\ref{mr})) offers a rapid and reasonably
accurate method of estimating the conductivity of plasma mixtures from
the pure plasma conductivities.  Another point worth noting is that
the mixing rule predicts that the larger the charge asymetry, the smaller
the amount of the more highly charged ion is needed to have a significant
effect on the mixture conductivity.

The second application of the model is to evaluate the Thomas-Fermi model
through comparison with the Kohn-Sham version.
Note that in both the TF and KS versions the cross section is evalutated
quantum mechanically using equation (\ref{trq}), only the generation
of $V^{MF}_{ie}$ changes.
In figure \ref{fig_compare} we compare these for a CH mixture and
for pure hydrogen and carbon plasmas.  The agreement between the
TF and KS versions improves for lower density and higher
temperatures, i.e. for lower degeneracy plasmas.  Since the
derivative of the Fermi-Dirac function is broader for
lower degeneracy, the energy integral in equation (\ref{sdc})
has a wider range of energies that are significant,
including higher energies where the TF cross section is a better approximation to KS.
The result will be therefore less sensitive to the details of the
relaxation time, hence use of the TF $V^{MF}(r)$ should be more
reasonable.  We note also that the TF model is just as accurate
for the mixture as it is for the pure hydrogen and carbon
plasmas (compared to the KS model).

\section{Conclusions\label{sec_con}}
A model for calculating the conductivity of dense plasma mixtures
has been presented.  The model is an extension to multicomponent
ionic mixtures of the model presented in reference \cite{starrett17},
and builds on the multicomponent electronic and ionic structure
model presented in reference \cite{starrett14b}.  The new conductivity model
is computationally efficient, taking a few minutes per density and temperature
point.  It can also reach temperatures much higher than the Fermi energy, in contrast
to other methods based on DFT \cite{hanson11,lambert11, desjarlais17}.

We have evaluated the new model by comparing it to DFT-MD simulations
that use the Kubo-Greenwood approximation.  The model is in reasonably good agreement
with these less approximate results for a variety of physical plasmas.

The conductivity of asymmetric mixtures (high-Z, low-Z) was investigated
with a newly proposed mixing rule.
This model predicts that the ion charge asymmetry is what drives the change
in the conductivity, and that the higher the charge asymmetry, the less of
the highly charged ion is needed to have a significant effect.
The mixing rule is expected to work best for high temperatures, and represents
a simple and rapid way to obtain reasonably accurate mixture conductivities from
the pure plasma conductivities and average ionizations.

Finally, it was found that using a Thomas-Fermi model for the potential of mean force
gave good agreement with the Kohn-Sham version for sufficiently high temperature or low density.

\section*{Acknowledgments}
This work was performed under the auspices of the United States Department of Energy under contract 89233218CNA000001.

\appendix
\section{Mixing rule \label{mix}}
In this appendix we show how we arrive at the new mixing rule, equation (\ref{mr}), which
aims to estimate the mixture conductivity from the pure plasma conductivities of the components,
at the same mass density and temperature.
Following reference \cite{lee84}, the momentum transport cross section can 
be approximated in terms of a Coulomb logarithm ($\log\Lambda$)
\begin{equation}
\sigma_{i,\mathrm{tr}}(\epsilon) = \frac{4 \pi e^4 \bar{Z}_i^2 \log\Lambda_i}{m^2 v^4}
\label{trq_lm}
\end{equation}
Making the ansatz
\begin{equation}
\sigma_{DC} = \sum\limits_{i=1}^{N} Y_i \, \sigma_{i, DC} 
\end{equation}
where $Y_i$ are coefficients, and $\sigma_{i, DC}$ are the conductivities
for pure plasmas of species $i$ only (at the same mass density and temperature as the mixture),
and using equation (\ref{sdc}) and (\ref{rt}), we have
\begin{equation}
1 = \sum\limits_{i=1}^{N} Y_i \, \frac{\sum\limits_{j=1}^{N} n_j^0 {\bar{Z}_j}^2 \log\Lambda_j F(\mu)}
{n_i^{0,p} {(\bar{Z}_i^p)}^2 \log\Lambda_i^p F(\mu_i^p)}
\label{sum_one}
\end{equation}
On the top line, the quantities with subscript $j$ refer to properties of ions in the mixture,
and on the bottom line, quantites with a superscript $p$ refer to pure plasma
properties, and
\begin{equation}
F(\mu) = \frac{2}{\pi^3} \int_0^\infty d\epsilon \, \epsilon^3 f(\mu)
\end{equation}

Equation (\ref{sum_one}) is satisfied if we choose
\begin{equation}
Y_i = x_i
\frac{n_i^{0,p} {(\bar{Z}_i^p)}^2 \log\Lambda_i^p F(\mu_i^p)}
{\sum\limits_{j=1}^{N} n_j^0 {\bar{Z}_j}^2 \log\Lambda_j F(\mu)}
\label{y1}
\end{equation}
which we can approximate as
\begin{equation}
Y_i \approx x_i
\frac{{(\bar{Z}_i^p)}^2 }
{\sum\limits_{j=1}^{N} x_j {(\bar{Z}_j^p)}^2 }
\end{equation}
This assumes that
\begin{equation}
\frac{n_j^0}{n_i^{0,p}} \approx
\frac{n_j^0}{n_I^{0}} = x_j,
\end{equation}
\begin{equation}
\frac{\log\Lambda_j F(\mu)}
{\log\Lambda_i^p F(\mu_i^p)} \approx
1
\end{equation}
and finally
\begin{equation}
\bar{Z}_i  \approx \bar{Z}_i^p 
\end{equation}
%Equation (\ref{ass_1}) will be accurate when $x_i$ is close to $1$ or when the
%ion accupy similar volumes.   

\section{DFT-MD simulations\label{dft}}
We have performed DFT-MD calculations with the Vienna ab-initio 
simulation package (VASP~\cite{kresse93, kresse94, kresse96, perdew96}), 
using the Generalized Gradient 
Approximation -- Perdew, Burke, Ernzerhof for the XC functional (GGA-PBE~\cite{perdew96, perdew97}).
We employed the GW – 3e$^-$ plane augmented wave (PAW) pseudopotential (PP) for Al, 
and the 4e$^-$ PAW PP for C~\cite{blochl94, kresse99}, with a planewave cutoff energy of 750 eV. The 
DFT-MD simulations were performed at constant temperature using the Nose-Hoover thermostat 
and a timestep of 1 fs. 

We followed the procedure validated for pure Al in the WDM regime, and discussed in 
a recent publication \cite{starrett19}. Pre-molten samples containing 64 atoms at density $\rho$ were first 
equilibrated at temperature T for at least 2 ps. From this run, ten snapshots separated 
by 50 fs were used to calculate the DC conductivity via optical analysis. The MD 
stage calculations were performed at the Gamma-point, while a 2x2x2 k-point mesh 
(generating 4 independent k-points) was used for the optical analysis. We imposed 
a number of bands $N$ during the MD such that the maximum occupation of the highest 
band is less than $1\times 10^{-4}$, and used $2N$ bands for the optical analysis. 

Regarding the C-H calculations, which are extensions of the calculations of reference 
\cite{starrett12a}, we also used PBE-PAW PP, with a cutoff energy 
of 700 eV. Simulation cells containing 128 to 250 atoms were tested, yielding results differing 
by at most 10\% (for the C-90\%/H-10\% case). All calculations were performed at the $\Gamma$-point.

\setlength{\tabcolsep}{0.5em} % for the horizontal padding
{\renewcommand{\arraystretch}{2.2}% for the vertical padding

\begin{table}[]
\caption{DFT-MD $\sigma_{DC}$ results for C$_{x}$--Al$_{1-x}$ at 10 g/cm$^3$ in units of 10$^6$ $\Omega^{-1}$ m$^{-1}$.  
\label{tab_alc}}
\begin{center}
\begin{tabular}{c c | c c }
& &  \multicolumn{2}{c} {Temperature [eV]} \\
& &  5 &  10 \\
\hline
\multirow{5}{*}{\rotatebox[origin=r]{90}{Carbon fraction, $x$}} & 0    & 1.473 &  1.227  \\
& 0.25 & 1.1   &  0.998  \\
& 0.5  & 1.042 &  1.047  \\
& 0.75 & 1.267 &  1.254  \\
& 1    & 1.719 &  1.661 
\end{tabular}
\end{center}
\end{table}

\begin{table}[]
\caption{DFT-MD $\sigma_{DC}$ results for C$_{x}$--H$_{1-x}$ at 10 g/cm$^3$ in units of 10$^6$ $\Omega^-1$ m$^{-1}$.  
\label{tab_ch}}
\begin{center}
\begin{tabular}{c c |  c }
& &  Temperature [eV] \\
& &  10 \\
\hline
\multirow{4}{*}{\rotatebox[origin=r]{90}{Carbon fraction, $x$}} & 0.1    & 4.4     \\
& 0.2    & 2.85    \\
& 0.4375 & 1.89    \\
& 0.5    & 1.8    
\end{tabular}
\end{center}
\end{table}

\bibliographystyle{unsrt}
\bibliography{phys_bib}

\end{document}